\documentclass[twocolumn,showpacs,prb,aps,floatfix]{revtex4}
\usepackage{graphicx}% Include figure files
\usepackage{dcolumn}% Align table columns on decimal point
\usepackage{bm}% bold math

\begin{document}

\title{Evidence for a charge Kondo effect in Pb$_{1-x}$Tl$_x$Te from
    measurements of thermoelectric power}% Force line breaks with \\

\author{M. Matusiak$^{1,2}$, E.M. Tunnicliffe$^1$, J.R. Cooper$^1$, Y. Matsushita$^{3,4}$,
and I.R. Fisher$^{3,5}$} \affiliation{$^1$Cavendish Laboratory, Department of Physics,
University of Cambridge, J.J. Thomson Avenue, Cambridge CB3 0HE, UK\\ $^2$Institute of
Low Temperature and Structure Research, Polish Academy of Sciences, P.O. Box 1410,
50-950 Wroclaw, Poland\\ $^3$Department of Applied Physics and Geballe Laboratory for
Advanced Materials,
 Stanford University, Stanford, California 94305-4035
\\ $^4$ Department of Materials Science and Engineering and Geballe Laboratory for Advanced
Materials, Stanford University,
\\ $^5$Stanford Institute for Materials and Energy Science,
SLAC National Accelerator Center, 2575 Sand Hill Road, Menlo Park, CA 94025, USA}

\date{\today}% It is always \today, today,
             %  but any date may be explicitly specified

\begin{abstract}
We report measurements of the thermoelectric power (TEP) for a series of
Pb$_{1-x}$Tl$_x$Te crystals with $x$ = 0.0 to 1.3$\%$. Although the TEP is very large
for $x$ = 0.0, using a single band analysis based on older work for dilute magnetic
alloys we do find evidence for a Kondo contribution of 11 - 18 $\mu V/K$. This
analysis suggests that $T_K$ is $\simeq$ 50 - 70 K,  a factor 10 higher than
previously thought.
\end{abstract}

\pacs{72.15.Qm, 72.15.Jf, 74.70.Ad}

 \maketitle

%\section{Introduction}
The traditional ``spin'' Kondo effect occurs in a non-magnetic host metal containing a
small concentration of magnetic impurities. Here the anti-ferromagnetic exchange
interaction ($J$) between the local magnetic moment and the
 conduction electrons
  gives rise
to a $-|J|^3 \log T$ term in the temperature ($T$) dependent electrical resistivity
$\rho(T)$  and other unusual properties \cite{Kondo,KondoReview,Hewson} including an
anomalously large and \textit{T}-dependent thermoelectric power (TEP or \textit{S}).
Below the Kondo temperature ($T_K$) which can be extremely small, the spin of the
impurity is compensated by a cloud of conduction electron spins extending over a
distance $\sim\hbar v_F/k_BT_K$, where $v_F$ is the Fermi velocity of the host metal.
There is a ``triple peak'' in the impurity density of states (DOS) \cite{Hewson}, with
two side lobes derived from
 the spin-split virtual bound state and a narrow peak of width $k_BT_K$ at the Fermi
energy.  It is thought \cite{Dzero} that a similar description applies to the charge
Kondo effect discussed here.

 Experimental \cite{MatsushitaPRL} and theoretical \cite{Dzero}  evidence has been reported for a charge Kondo effect in Pb$_{1-x}$Tl$_x$Te crystals
   with $x$ $\geq$ 0.3$\%$. It arises because  Tl$^+$ and Tl$^{3+}$ ions both have filled shells, i.e. 6s$^2$
    and 6s${^0}$,
  that can be  more stable than the 6s$^1$ state of Tl$^{2+}$.
   So when Tl is in an environment favorable for divalency there can be two degenerate,
    or nearly degenerate, non-magnetic charge states Tl$^+$ and Tl$^{3+}$ that can be
    described by the Anderson model and can also
     give rise to a Kondo effect \cite{Dzero}.  The resulting charge fluctuations are
     thought to be important for superconductivity \cite{Dzero, MatsushitaPRL,MatsushitaPRB} in Pb$_{1-x}$Tl$_x$Te.
       In the present paper we report TEP data for single
     crystals of
      Pb$_{1-x}$Tl$_x$Te  and
   discuss evidence for an anomalous  contribution associated with a charge Kondo
      effect.

 %\section{Experimental Details}
The crystals of  Pb$_{1-x}$Tl$_x$Te with $x$ =0, 0.2, 0.3, 0.6, 1.1 and 1.3$\%$ were
from the same preparation batches studied previously
\cite{MatsushitaPRL,MatsushitaPRB,MatsushitaPhD}.
 Their thermopower was measured using CuBe wires \cite{LakeShore},
whose TEP was measured separately relative to a cuprate superconductor and found to be
small with a maximum  of 0.2 $\mu V/K$ near 30 K. Initially the temperature
 gradient, $\Delta T$, was measured  using a constantan-chromel
 thermocouple made of 25 $\mu m$ diameter wires, and glued to the $mm$-sized crystals
  with GE varnish, but for the data reported here very small diode thermometers
  \cite{LakeShore} were used.
  These were less straightforward
   to mount, but gave more reliable measurements of $\Delta T$, especially at
    low $T$ where the sensitivity of the thermocouple decreases.  The
     diodes were also attached to the sample using GE varnish, and the distance
      between them, measured to $\pm 3$ to 5 $\%$ using a binocular microscope,
   is the main source of error in the TEP.

%\section{Results and Analysis}
For many dilute metallic alloys $\rho(T)$ approximately obeys Matthiessen's rule which
states that the contributions $\rho_j$ arising from two or more different scattering
mechanisms ($j$ = 0,1,2..) are simply additive.   It generally holds reasonably well
when $T$ is comparable to the Debye temperature ($\Theta_D$) or when the impurity
resistivity is smaller than that caused by electron-phonon scattering, but at lower
$T$ there can be significant deviations.  One probable reason is that conservation of
crystal momentum (\textbf{k}) is less strict in an alloy because  the finite electron
mean free path limits the size of electron wave packets \cite{CaplinMR}. The
equivalent Nordheim-Gorter (NG) rule \cite{MacDonald62}
 for combining contributions
$S^{d}_{j}$ to the total electron diffusion TEP $S^d$ from different scattering
mechanisms in the same band is:

\begin{equation}
\\ S^{d}  = \frac{\Sigma_{j}S^{d}_{j} \rho_{j}}{\Sigma_{j}\rho_{j}}
 \label{TEP1}
\end{equation}

It relies on  Matthiessen's rule being obeyed and on the scattering  being effectively
elastic, which is only true for $T\gtrsim \Theta_D$ or in the residual resistivity
($\rho_{res}$) region.  Resistivity  data \cite{MatsushitaPhD} for  Pb$_{1-x}$Tl$_x$Te
crystals from the same preparation batches are shown in Fig. 1. Matthiessen's rule is
obeyed reasonably well for $x\geq0.6\%$ for all $T$ and for $x=0.3\%$ for $T<150$ K.
The insert shows that $\rho_{res}$ is small for $x \leq 0.2 \%$ but then increases
linearly  with $x$ for $x \geq 0.3\%$. This is consistent with other evidence that
valency skipping sets in near $x = 0.3\%$ \cite{MatsushitaPRL} and, together with
Matthiessen's rule, implies that there are no gross changes in electronic structure at
higher $x$.  The linearity also suggests
 that the Tl impurities act as independent scattering centers for $x\geq 0.3\%$.
TEP data measured for the six samples are shown in Fig. 2.  All Tl-doped samples have
$S\sim$ 100 - 140 $\mu V/K$ at 300 K with similar slopes  and  a positive
curvature at lower $T$, while the pure PbTe sample has a larger TEP, 300 $\mu V/K$ at
300 K, and  a negative curvature.

\begin{figure}[hbtp]
\begin{center}
\includegraphics[width=80mm,height=80mm]{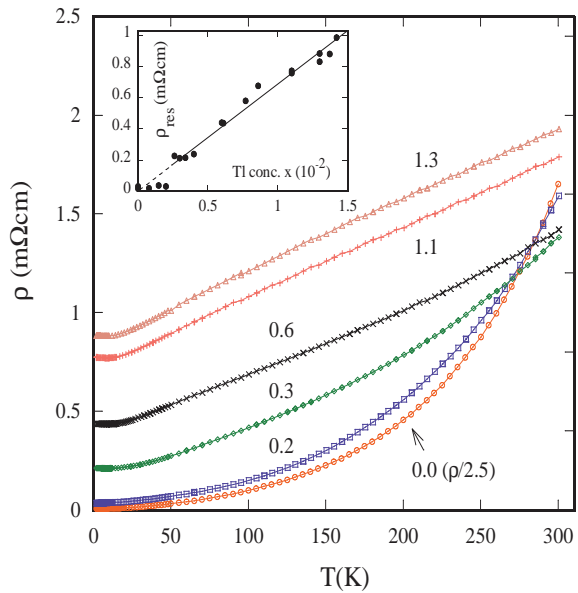}
\caption{Color online: resistivity \textit{vs.} temperature for Pb$_{1-x}$Tl$_x$Te crystals \cite{MatsushitaPhD} from
 the
same preparation batches as those studied here, values of $x$ given in $\%$. The insert shows residual
 resistivity \textit{vs.}
 $x$
for all crystals measured, $\rho(T)$ data for all $x$ values in the insert are given
in Ref. \onlinecite{MatsushitaPhD}.}
\end{center}
\end{figure}

 Analysis of various Kondo
alloys (e.g. \underline{Au}Fe, \underline{Au}Mn and
        \underline{Au}Cr) \cite{CooperJmag,CooperFord}
 suggests  that  $S^d$ in Eqn. 1  contains three terms, $S_0$, $S_1$ and $S_K$. $S_0$ arises
 from the energy
dependence of the potential scattering  term $V$ in the Kondo Hamiltonian, and is
 related to the energies and widths of the 3d virtual bound states. It is linear in
$T$ and in Eqn. 1 is weighted by $\rho_{res} \equiv \rho_0$.  The electron-phonon
scattering term, $S_1$, is also expected to be linear in $T$ both for $T \gtrsim
\Theta_D/5$ and $T < \Theta_D/10$  but with a factor of 3 smaller slope at low $T$.
For PbTe, heat capacity data \cite{MatsushitaPRB} give $\Theta _D$ =168 K.
 $S_0$ , $S_1$,
$\rho_0$ and  $\rho_1$ are often calculated using first order perturbation theory,
while  $S_K$, the Kondo TEP contribution, arises from  higher order scattering
processes involving  non-cancelling Fermi factors. It is also weighted by $\rho_0$ in
Eqn. 1.  A broad peak in $S^d$ is often observed near the Kondo temperature $T_K$, and
often used to estimate $T_K$ \cite{Hewson,Heeger}. However it has been argued that
$S_K$ could  be constant for $T\gtrsim T_K$, \cite{CooperVucic,CooperJmag}, as
predicted by high $T$ perturbation theory \cite{KondoReview} and that the fall above
the peak is  caused by the other $T$-dependent terms in Eqn. 1. For $T \lesssim
0.1-0.15 T_K$ \cite{CooperVucic}, $S_K$ falls to zero as $T^1$, and fits $S_K
=AT/(T+0.35T_K)$ for $T\lesssim T_K$.

\begin{figure}[hbtp]
\begin{center}
\includegraphics[width=80mm,height=80mm]{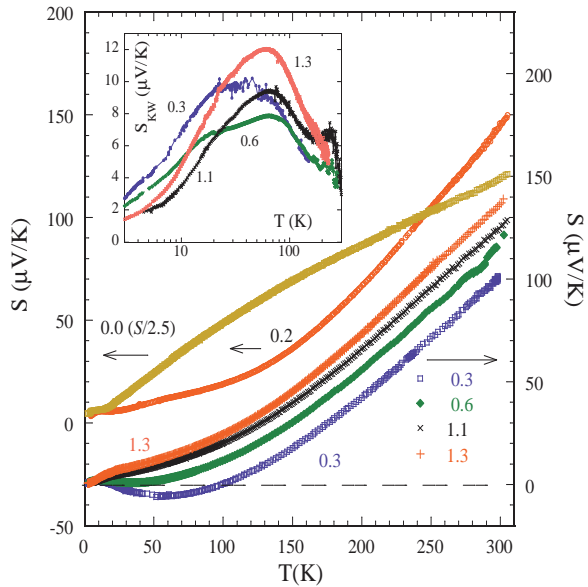}
\caption{Color online: measured values of the TEP for Pb$_{1-x}$Tl$_x$Te crystals with
$x$ = 0.0 and 0.2$\%$ -left hand scale and $x$ = 0.3, 0.6, 1.1 and 1.3 $\%$ - right
hand scale. The inset shows the weighted Kondo contribution, $S_{KW} \equiv
S_K\rho_0/\rho(T)$ for the values of $x(\%)$ shown.}
\end{center}
\end{figure}
In view of the above discussion, we expect
  $S_K$  to be constant or to fall
slowly with $T$ above a  cut-off temperature $T_0> T_K$, and so above $T_0$ Eqn. 1 gives:

\begin{equation}
\\ S  = \frac{\alpha T \rho _0  + \beta T \rho _1 +\gamma \rho _0}{\rho_0+\rho_1(T)}
   \label{TEP3}
\end{equation}
where  $\alpha$, $\beta$ and $\gamma$ are independent of $x$ and $T$. Fits to Eqn. 2
were made from 300 K to $T_0$ = 90,
70, 50 or 30 K, with $\alpha$, $\beta$ and $\gamma$ as free parameters and $\rho_0$
 and $\rho_1(T)$ obtained by fitting the data in Fig. 1 to $\rho = A + BT + CT^2$. Values of
 $\alpha$, $\beta$ and $\gamma$ and $A$, $B$ and $C$ are given in Table 1.
  $T_0$ =70 K gave slightly better fits. Fits  from 150  to 70 K gave similar results and
  for $x= 0.3\%$ we used this range because
    of the curvature in
    $\rho(T)$ shown in Fig.1 for this sample.  Below 70 K, $S_K(T)$  was
    obtained from  Eqn. 2
     by using the formula $S_K(T) = S_{meas}\ \rho/\rho_0 -\alpha T - \beta T\rho_1/\rho_0$
  and is shown in the main part of Fig. 3a. Our fitting procedure
  should give a constant value for
  $S_K$ above 70 K. The variations seen in Fig. 3a   arise from residual errors in the
  fits multiplied up
   by $\rho/\rho_0$ and are only a few
  times the noise level.

  For an isotropic  parabolic band with Fermi energy $E_F$
   $\beta \simeq \pi^2k{_B^2}/(eE_F)$ for $T \gtrsim
\Theta_D/5$ and  $\pi^2k{_B^2}/3E_F$ for  $T \leq\Theta_D/10$
\cite{MacDonald62,Ziman}, i.e. $T \leq$ 17 K for PbTe. The  $\beta$ values in Table I
give $E_F$ between 73 to 160 $meV$ so their sign and magnitude are reasonably
consistent
 with $E_F$  being measured
relative to the bottom of  the $\Sigma$ band \cite{MatsushitaPhD}, despite the fact
that the hole pockets are actually ellipsoidal.  But the changes of $\beta$ with $x$
are not understood since Hall data \cite{MatsushitaPRB} suggest a slight increase in
hole concentration, i. e. in $E_F$ with $x$. The magnitudes of $\alpha$ are similar to
$\beta$ but of opposite sign corresponding to higher energy holes being more strongly
scattered by the Tl impurities. Within a Kondo picture  this arises from the asymmetry
of the Friedel-Anderson virtual bound states describing the valency fluctuations, but
a low value of $E_F$ for the $\Sigma$ band could also affect $\alpha$. Finally the
values of $\gamma$ ranging from 11 to 18 $\mu V/K$ are similar to those associated
with the traditional Kondo effect and the asymmetric Anderson model.

\begin{table}
\caption{Fitting parameters, $A$, $B$, $C$, for $\rho(T)$ and  $\alpha$, $\beta$,
$\gamma$, for TEP. }
\begin{ruledtabular}
\begin{tabular}{l|c|c|c|c|c|c}
$ $&$A $&$B $&$C $&$ \alpha $&$ \beta $&$ \gamma $\\
$x\%$ &$ m\Omega cm$&$ m\Omega cm/K$&$m\Omega cm/K^2$&$\mu V/K^2 $&$\mu V/K^2 $&$\mu V/K $\\
\hline $0.3$&$0.2017$&$1.080\textrm{x}10^{-3}$&$9.29\textrm{x}10^{-6}$&$-0.581$&$0.446$&$14.3$\\
\hline $0.6$&$0.4124$&$2.315\textrm{x}10^{-3}$&$3.37\textrm{x}10^{-6}$&$-0.362$&$0.682$&$11.5$\\
\hline $1.1$&$0.7371$&$3.343\textrm{x}10^{-3}$&$5.97\textrm{x}10^{-7}$&$-0.211$&$0.837$&$12.1$\\
\hline $1.3$&$0.8472$&$3.640\textrm{x}10^{-3}$&$-2.6\textrm{x}10^{-8}$&$-0.211$&$0.947$&$17.7$\\
\end{tabular}
\end{ruledtabular}
\label{summary}
\end{table}

\begin{figure}[hbtp]
\begin{center}
\includegraphics[width=70mm,height=95mm]{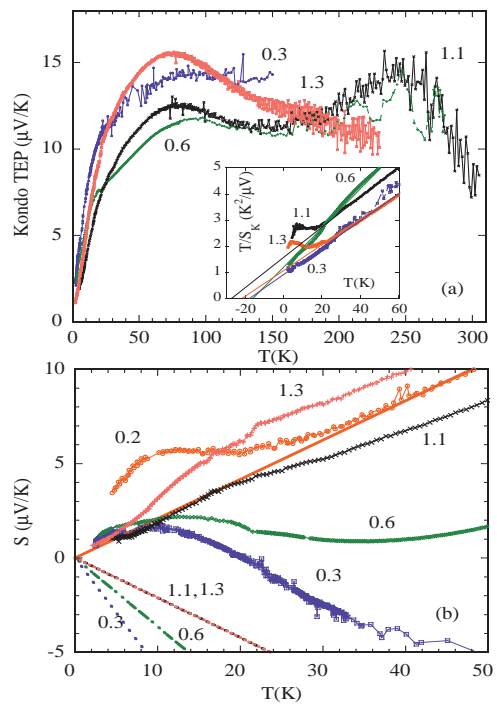}
\caption{Color online: (a) Kondo contribution to the TEP of Pb$_{1-x}$Tl$_x$Te single
crystals with values of $x$ shown in $\%$. The insert shows  $T/S_K$ vs. $T$ plots
used to find
 $T_K$ from the empirical, low $T$, law \cite{CooperVucic}, $S_K
=AT/(T+0.35T_K)$. (b) Raw TEP data for the values of $x$ in $\%$ shown. A solid line
shows $S^d$ for $x=0.2\%$ and dashed lines show $\alpha T$ for  $x\geq0.3\%$}
\end{center}
\end{figure}
 Fits to Eqn. 2 for the
    $x$ = 0.0 and 0.2 $\%$ samples were not good and are not shown. As shown in Fig. 1
    for these
     two samples $\rho_{res}$ values  are very low
     and  $\rho(T)$ much more curved than for $x\geq 0.3\%$. These
      differences and Hall data \cite{MatsushitaPRB} suggest that
    the $L$ band dominates electronic transport
 for low $x$
         while for   $x\geq 0.3\%$
      the  $\Sigma$ band plays a leading role \cite{MatsushitaPRL,MatsushitaPhD}.  The TEP of the
    0.2 $\%$
    crystal varies  as $AT$ between 40 and 100 K with $A$ = 0.2 $\mu V/K^2$ but rises sharply at higher
    $T$. This value of $A$ is reasonably  compatible with $E_F$ = 200 $meV$ obtained from analysis of Hall
data \cite{MatsushitaPRB} but the strong increase at higher $T$ is not understood.
    The deviations from linearity for $x$ = 0.2 $\%$ shown in Fig. 3b   are ascribed to the gradual
    onset of phonon drag
      below 40 K. This  is considerably lower than for  most
        metals where it is usually $\simeq \Theta_D$ \cite{MacDonald62} and  may arise because  both the
         $L$ and $\Sigma$ hole pockets have small  Fermi
     wave-vectors ($k_F$).  So electron-phonon scattering
     within a pocket will be suppressed
       to lower $T$ until typical phonon
      wave-vectors $\sim (T/\Theta_D)(\pi/a)$, where $a$ is the lattice spacing,
     become smaller than $2k_F$. We cannot  assume that  phonon drag corrections
        for the $\Sigma$ band, i.e. $x\geq0.3\%$,
     will be  similar to those for $x=0.2\%$.
       Perhaps the only way
      to obtain the phonon drag term for $x\geq0.3\%$ is to suppress superconductivity by applying a magnetic field
and measure to much lower $T$ where  it will eventually vary as $T^3$.
      Below T $\sim$ 8 K the raw data in Fig. 3b show an approximately $T$-linear, and
    $x$-independent decrease that is typical of the Kondo  effect
     for $T\lesssim$ 0.1 - 0.2 $T_K$ \cite{CooperVucic}.  However according to the preceding analysis the $\alpha T$ terms
      shown by the dashed lines in Fig. 3b are still present at low $T$.
      They should be subtracted to obtain $S_K(T)$ and this spoils the $x$-independence to
       some extent.

 We
 estimated $T_K$ using two methods.  The first one is based on the widely held view
 \cite{Heeger,Hewson} that
 that there is an $x$-independent peak in the TEP at $T_K$ in dilute magnetic alloys, where the host metal has a very
 small TEP. If the TEP of the PbTe host metal and
   $S_0$ were both
small then
  we would measure a weighted value $S_{KW} \equiv S_K \rho_0/\rho$. Plots
 of $S_{KW}$ found in this way, using the $S_K(T)$ data in Fig. 3 are shown
  in the insert to Fig. 2. We see that
 there are indeed broad peaks near 60 K for $x = 0.6$, $1.1$ and $1.3\%$.
 The second method is based on the empirical
law $S_K = AT/(T+0.35T_K)$ \cite{CooperVucic} for which plots of $T/S_K$ vs. $T$  give
a straight line extrapolating to $S_K$ = 0 at  -0.35$T_K$.  As shown in the insert to
 Fig. 3a such plots are reasonably linear
and give $T_K$ values ranging from 45 to 75 K for the 4 samples. The deviations
below 20 K for $x=1.1$ and $1.3\%$ could arise from phonon drag.
Values of $T_K$ obtained in these two ways are significantly higher than
 $T_K \sim$ 6 K \cite{MatsushitaPRL} which was estimated
 by fitting
  resistivity data to $\rho = \rho(0)[1-(T/T_K)^2]$ but with
   considerable uncertainty in the appropriate value of $\rho(0)$.
  On the other hand, using the fact that the $T^2$
   law
  normally extends  up to 0.1 $T_K$ \cite{RizzutoReview},
gives $T_K \gtrsim$ 40 K - in closer agreement with our estimates from the TEP.

Although we should be cautious about applying spin Kondo formulae to  a doped
semiconductor, PbTe, the observation of
 Matthiessen's rule is consistent with a single-band metallic picture and minor changes
 in hole concentration above
 $x$ = 0.3$\%$.
Taking    $T_K$ = 60 K \cite{FootnoteNG} does affect some of the previous conclusions
\cite{MatsushitaPRL,MatsushitaPRB}. For example on the basis of the measured specific
heat coefficient $\gamma(0)$ and the depth of the resistivity minimum it was suggested
\cite{MatsushitaPRL,MatsushitaPRB} that only $x_{eff}/x \sim 0.01$ of the Tl
impurities were degenerate to within $T_K$ = 6 K and hence only these were effective
Kondo scatterers. With $T_K$ = 60 K, $x_{eff}/x $ becomes $\sim 0.1$ and also the
radius of the charge cloud is smaller, $\sim$ 25 nm, taking $k_ F$ = 10$^7$ cm$^{-1}$
and $m = 0.6 m_e$, i.e. $v_F$ = 2 10$^7$ cm/sec. Accordingly at $x$ = 1$\%$, on
average there would still be   about 10 other Tl atoms within the charge cloud of one
impurity, but because $x_{eff}/x \sim 0.1$ only one of these would be a Kondo
scatterer. We note that the charge clouds must overlap and be delocalized at low $T$
(as are the Kondo states in heavy Fermion superconductors) because otherwise they
would not contribute
 to superconductivity and there would be a residual $\gamma(0)$ in the
superconducting state. Such considerations of charge overlap may make detailed
analysis in terms of the $L$ and $\Sigma$ bands of undoped PbTe less straightforward
than previously thought.

Assuming that there is also a narrow peak $\sim k_BT_K$ wide in the  DOS for  charge
Kondo impurities,  then the field\textit{} scale for magneto-transport  effects is
larger $\sim k_B T_K/\mu _B$.  Also one would expect a magnetic susceptibility
contribution $\chi(T)= \mu_B^2x_{eff}N_{AV}/k_B(T+T_K)$ emu/mole where $N_{AV}$ is
Avogadro's number, $\mu_B$, the Bohr magneton, and from above $x_{eff }\sim 0.1x$.
This would be undetectable in the available $\chi(T)$ data
\cite{MatsushitaPRL,MatsushitaPhD} but might show up in detailed measurements and
analysis, like those made for AlMn alloys \cite{Cooper&Miljak}.  There are some
indications that $x_{eff}/x$ could be larger than 0.1.  Using the formula
\cite{Hewson} $\gamma(0) = 0.4128\pi^2k_B/(6T_K)$ gives $\gamma(0)$ = 0.94
mj/mole/K$^2$ for $x$ = 1$\%$ compared with the average slope from the \emph{raw}
experimental data \cite{MatsushitaPRB} of $\gamma(0)/x$ = 0.57 mj/mole/K$^2$/$\%$
implying that $x_{eff}/x\simeq$ 0.6. We also note that the measured values of $\Delta
\chi(0)/\Delta \gamma(0)$ are remarkably close to $6\mu_B^2/\pi^2k_B^2$, the value
expected from spin Kondo theory, namely a Wilson ratio of 2 rather than 1 for
non-interacting fermions. However for $x_{eff}/x\simeq$ 0.6 it is difficult to
understand the $T$ dependence since the $\chi(T)$ curves for $x$ between 0.3 and 1.3
$\%$ are essentially parallel. Although the effective mass and band gap of PbTe are
$T$ dependent \cite{Lashkarev}, it seems unlikely that this would exactly compensate
any $1/(T+T_K)$ behavior from the Kondo effect.

In summary, the large TEP of the host material makes the data analysis  less
straightforward than for noble metal-based Kondo alloys.  Using a provisional
single-band picture we do see reasonably clear evidence for a Kondo effect but with
$T_K$ a factor of 10 or so larger than the earlier work.  Band calculations of the TEP
of PbTe as a function of hole doping might  provide a further test of this conclusion.

We are grateful to T.H. Geballe, J. Schmalian and V. Zlati\'{c} for their helpful
comments. Work at Stanford was supported by the U.S. DOE, Office of Basic Energy
Sciences under contract DE-AC02-76SF00515 and that in Cambridge by the U.K. EPSRC.
%\section{Summary and Conclusions}

\end{document}